\documentclass[conference]{IEEEtran}
\IEEEoverridecommandlockouts

\usepackage{cite}
\usepackage{comment}
\usepackage{float}
\usepackage{amsmath,amssymb,amsfonts,amsthm, bbm}
\usepackage{algorithm}
\usepackage{algpseudocode}
\usepackage{graphicx}
\usepackage{subcaption}
\usepackage{textcomp}
\usepackage{xcolor}
\usepackage{tikz}
\usepackage{mathtools}
\usepackage[normalem]{ulem}
\usepackage[margin=0.75in]{geometry}
\usepackage{scalerel}
\usepackage{placeins}
\usepackage{overpic}
\usepackage[colorlinks=true]{hyperref}

\begin{document}

\newtheorem{hypotheses}{Hypotheses}
\newtheorem{problem}{Problem}
\newtheorem{example}{Example}
\newtheorem{definition}{Definition}
\newtheorem{assumption}{Assumption}
\newtheorem{theorem}{Theorem}
\newtheorem{lemma}{Lemma}
\newtheorem{corollary}{Corollary}[theorem]
\newtheorem{proposition}{Proposition}
\newtheorem*{remark}{Remark}
\newtheorem{conjecture}{Conjecture}
\numberwithin{assumption}{section}

\newcommand{\edit}[1]{\textcolor{blue}{#1}}
\newcommand{\ignore}[1]{}
\newcommand{\diff}{\mathop{}\!\mathrm{d}}

\algnewcommand{\Inputs}[1]{%
  \State \textbf{Inputs:}
  \Statex \hspace*{\algorithmicindent}\parbox[t]{.8\linewidth}{\raggedright #1}
}
\algnewcommand{\Initialize}[1]{%
  \State \textbf{Initialize:}
  \Statex \hspace*{\algorithmicindent}\parbox[t]{.8\linewidth}{\raggedright #1}
}

\title{\LARGE \bf Safe Reference Tracking and Collision Avoidance for Taxiing Aircraft Using an MPC-CBF Framework}

\author{Brooks A. Butler, Zarif Cabrera, Andy Nguyen, and Philip E. Par\'{e}$^*$
\thanks{*Brooks A. Butler is with the Department of Electrical Engineering and Computer Science at the University of California, Irvine, Zarif Cabrera is with the Department of Electrical Engineering and Computer Science at Howard University, Andy Nguyen is with the Department of Mechanical Engineering at Augustana College, and Philip. E. Par\'e is with the Elmore Family School of Electrical and Computer Engineering at Purdue University. Emails: \{bbutler2@uci.edu, zarif.cabrera@bison.howard.edu, andynguyen21@augustana.edu, and philpare@purdue.edu\}.
This work was partially funded by grants from the National Science Foundation (NSF-ECCS \#2238388 and NSF- CNS \#1836900) as well as  a doctoral fellowship of the Purdue-Windracers Center on AI for Digital, Autonomous and Augmented Aviation (AIDA$^3$).} 
}

\maketitle

\begin{abstract}
In this paper, we develop a framework for the automatic taxiing of aircraft between hangar and take-off given a graph-based model of an airport. We implement a high-level path-planning algorithm that models taxiway intersections as nodes in an undirected graph, algorithmically constructs a directed graph according to the physical limitations of the aircraft, and finds the shortest valid taxi path through the directed graph using Dijkstra's algorithm. We then use this shortest path to construct a reference trajectory for the aircraft to follow that considers the turning capabilities of a given aircraft. Using high-order control barrier functions (HOCBFs), we construct safety conditions for multi-obstacle avoidance and safe reference tracking for simple 2D unicycle dynamics with acceleration control inputs. We then use these safety conditions to design an MPC-CBF framework that tracks the reference trajectory while adhering to the safety constraints.
We compare the performance of our MPC-CBF controller with a 
PID-CBF control method via simulations.
\end{abstract}

\section{Introduction}
Developing fully autonomous aircraft poses a variety of technically complex challenges. Underlying these challenges is an emphasis on the criticality of system operations, where any system failure or error may lead to catastrophic outcomes. Thus, in this work, we are interested in applying tools from control theory to autonomous aircraft that can provide much-needed assurance on system behavior. Specifically, we provide a framework for safe autonomous taxiing for aircraft from designated hangars to runways for takeoff (and from runways to hangars after landing) in airport environments.
Among the many challenges posed by autonomous taxiing, we focus on the following capabilities:
\begin{enumerate}
    \item Automatic route planning from a starting hangar to a takeoff runway (and vice versa) that adheres to the physical limitations of the given aircraft
    \item Safe reference path tracking 
    \item Collision avoidance of unplanned obstacles
\end{enumerate}

With the rise of autonomous vehicles in recent years, developing a framework for autonomous taxiing is an active area of research across controls and aviation engineering \cite{zhang2020software,lu2016improved,morris2015self,gaikwad2023developing,zaninotto2019design,fremont2020formal,sirigu2018autonomous,soltani2020eco}. Current approaches range from learning-based methods \cite{lu2016improved, gaikwad2023developing} to optimization \cite{zaninotto2019design} and correct-by-construction verification \cite{zhang2020software}. Additionally, the current work considers the means of aircraft taxiing at different levels of autonomy, including self-driving aircraft-towing vehicles \cite{morris2015self} and human-in-the-loop verification \cite{zaninotto2019design}. In this work, however, we focus on implementing an auto-taxiing framework for a fully autonomous aircraft that is fully self-driven (i.e., without the use of towing vehicles or human input).

The study of safety-critical control has seen a significant resurgence in recent years, due largely to the introduction and refinement of \textit{control barrier functions} (CBFs)~\cite{ames2016control, ames2019control}. Additionally, due to the ease with which CBF safety conditions can be implemented into optimization methods using quadratic programming, including CBFs in the construction of model-predictive control (MPC) frameworks has been a natural extension to the design of safe controllers for dynamic systems \cite{zeng2021safety,vangasse2023mpc,wang2024robust,otsuki2023hierarchical,jankovic2021collision}. Many of the proposed MPC-CBF frameworks focus on either obstacle avoidance \cite{zeng2021safety,wang2024robust,otsuki2023hierarchical} or inter-agent collision avoidance \cite{vangasse2023mpc,jankovic2021collision}; however, to the best of our knowledge, the developing an MPC-CBF framework for reference tracking, specifically for application to taxiing aircraft, remains an open problem.

Thus, in this paper, we make the following contributions toward developing a framework for fully autonomous taxiing aircraft:
\begin{itemize}
    \item  We implement multi-obstacle avoidance for self-taxiing aircraft using an MPC-CBF framework.
    \item  Using a graph-based path planning algorithm for taxiing aircraft from \cite{zhang2020software} to generate a reference trajectory, we implement a safety condition for safe reference tracking within the same MPC-CBF framework.
    \item We compare the performance of our MPC-CBF framework with a model-free PID-CBF method.
\end{itemize}
After defining the necessary notation, we introduce some preliminary concepts from safety-critical control in Section~\ref{sec:prelim}, then describe a path-planning algorithm for generating a viable reference trajectory in Section~\ref{sec:pathplanning}. We then define safety conditions for obstacle avoidance and safe reference tracking in Section~\ref{sec:safety_framework} and demonstrate our framework via simulations in Section~\ref{sec:simulations}.

\subsection{Notation}
Let $|\mathcal{C}|$ denote the cardinality of the set $\mathcal{C}$. $\mathbb{R}$ and $\mathbb{N}$ are the set of real numbers and positive integers, respectively. Let $D^r$ denote the set of functions $r$-times continuously differentiable in all arguments. 
A monotonically increasing continuous function $\alpha: \mathbb{R}_{+} \rightarrow \mathbb{R}_{+}$ with $\alpha(0) = 0$ is termed as class-$\mathcal{K}$. The dot product of two same-sized vectors $a$ and $b$ is notated as $a \cdot b$ and $\Vert a \Vert$ is the 2-norm of vector $a$.
We define $[n] \subset \mathbb{N}$ to be a set of indices $\{1, 2, \dots, n\}$.
We define the Lie derivative of the function $h:\mathbb{R}^N \rightarrow \mathbb{R}$ with respect to the vector field generated by $f:\mathbb{R}^N \rightarrow \mathbb{R}^N$ as
\begin{equation}
    \mathcal{L}_f h(x) = \frac{\partial h(x)}{\partial x} f(x).
\end{equation}
We define higher-order Lie derivatives with respect to the same vector field $f$ with a recursive formula \cite{robenack2008computation}, where $k>1$, as
\begin{equation}
    \mathcal{L}^k_f h(x) = \frac{\partial \mathcal{L}^{k-1}_f h(x)}{\partial x} f(x).
\end{equation}
We define a networked system using a graph $\mathcal{G} = (\mathcal{V}, \mathcal{E})$, where $\mathcal{V}$ is the set of $n = \vert \mathcal{V} \vert$ nodes, $ \mathcal{E} \subseteq \mathcal{V}\times \mathcal{V} $ is the set of edges. Let $\mathcal{N}_i$ be the set of all neighbors with an edge connection to node $i \in [n]$, where 
\begin{equation}
    \mathcal{N}_i = \{j \in [n]\setminus \{i\}: (i,j) \in \mathcal{E} \}.
\end{equation}

\section{Preliminaries} \label{sec:prelim}
In this section, we introduce some preliminary concepts from safety-critical control including control barrier functions \cite{ames2016control, ames2019control} and high-order control barrier functions \cite{xiao2021high,tan2021high}, which will be used to enforce our taxiing safety constraints.

\subsection{Control Barrier Functions}

Consider a dynamical system of control affine form
\begin{equation} \label{eq:dyn_gen}
    \dot{x} = f(x) + g(x)u,
\end{equation}
where $x \in \mathbb{R}^n$, $u \in \mathcal{U} \subset \mathbb{R}^m$, and $f$ and $g$ are locally Lipshitz continuous. Let the set $\mathcal{C} \subset \mathbb{R}^n$ include all safe states for the system, where safety in the context of this paper is defined as the taxiing aircraft avoiding collisions with all obstacles and always staying on the runway. The task is then to design a controller that guarantees the forward invariance of the set $\mathcal{C}, \forall t \geq 0$. If we can encode our safety constraints $\mathcal{C}$ through a function $h(x):\mathbb{R}^n \rightarrow \mathbb{R}$ such that
\begin{equation} \label{eq:safeset_gen}
    \mathcal{C} = \{ x \in \mathbb{R}^n: h(x) \geq 0 \},
\end{equation}
then we can evaluate if $h(x)$ meets the criteria for being a control barrier function with the following definition and lemma.

\begin{definition}
    Given a dynamical system \eqref{eq:dyn_gen} and a set $\mathcal{C} \subset \mathbb{R}^n$ defined by \eqref{eq:safeset_gen}, the function $h$ is a \textbf{control barrier function} (CBF) if there exists a class-$\mathcal{K}$ function $\alpha$ such that, for all $x \in \mathcal{C}$,
    \begin{equation*}
        \sup_{u \in \mathcal{U}} \left[\mathcal{L}_f h(x) + \mathcal{L}_g h(x) u + \alpha(h(x))\right] \geq 0.
    \end{equation*}
\end{definition}

\begin{lemma} \cite[Proposition 1]{ames2016control}
    Consider a CBF $h(x)$. Then, any locally Lipschitz controller $u(x): \mathbb{R}^n \rightarrow \mathbb{R}^m$ such that
    \begin{equation}
        u(x) \in \{\mathcal{L}_f h(x) + \mathcal{L}_g h(x) u + \alpha(h(x)) \geq 0\}
    \end{equation}
    will render the set $\mathcal{C}$ forward invariant for system \eqref{eq:dyn_gen}.
\end{lemma}

\subsection{High-Order Control Barrier Functions}
High-order barrier functions (HOBFs) \cite{xiao2021high,tan2021high} are an extension of barrier functions that are useful when considering systems with control inputs in a higher degree of the system dynamics.

\begin{definition}
    \cite[Least relative degree]{tan2021high}:
    Given an arbitrary set $\mathcal{C} \subset \mathbb{R}^n$, an $r$th-order differentiable function $h:\mathbb{R}^n \rightarrow \mathbb{R}$ has \textbf{least relative degree} $r$ in $\mathcal{C}$ for system \eqref{eq:dyn_gen} if $\mathcal{L}_g\mathcal{L}_f^k h(x) = \mathbf{0}, \forall x \in \mathcal{C}$ for $k=1, \dots, r-2$.
\end{definition}

Given an $r$th-order differentiable function $h$ and extended class-$\mathcal{K}$ functions $\alpha^1(\cdot), \dots, \alpha_i^k(\cdot)$, we define a series of functions in the following form
\begin{equation} \label{eq:HO_funcs}
    \begin{aligned}
        \psi^0(x) := h(x), \; \psi^k(x) := \dot{\psi}^{k-1}(x) + \alpha^k(\psi^{k-1}(x)),
    \end{aligned}
\end{equation}
for $1 \leq k \leq r$. These functions provide definitions for the corresponding series of sets
\begin{equation} \label{eq:HO_sets}
    \begin{aligned}
        \mathcal{C}^k &:= \{ x \in \mathbb{R}^n: \psi^{k-1}(x) \geq 0 \},
    \end{aligned}
\end{equation}
which yields the following definition and lemma.

\begin{definition}
    Consider a control system defined by \eqref{eq:dyn_gen} with least relative degree $r$ and an $r$th-order differentiable function $h:\mathbb{R}^n \rightarrow \mathbb{R}$. The function $h$ is a \textbf{high-order control barrier function} (HOCBF) if there exist differentiable class-$\mathcal{K}$ functions $\alpha^k, k \in \{1, \dots, r\}$ such that, for all $x \in \bigcap_{k=1}^r \mathcal{C}^k$,
    \begin{equation}
        \sup_{u \in \mathcal{U}} \left[\mathcal{L}_f \psi^{r-1}(x) + \mathcal{L}_g \psi^{r-1}(x) u + \alpha^r(\psi^{r-1}(x))\right] \geq 0.
    \end{equation}
\end{definition}

\begin{lemma} \label{lem:HOCBF} 
\cite[Theorem 1]{tan2021high}
    Consider an HOCBF $h, \psi^{k-1}, k \in \{1, \dots, r\}$, defined in \eqref{eq:HO_funcs}. Then, any locally Lipschitz controller $u(x): \mathbb{R}^n \rightarrow \mathbb{R}^m$ such that
    \begin{equation}
        u(x) \in \{\mathcal{L}_f \psi^{r-1}(x) + \mathcal{L}_g \psi^{r-1}(x) u + \alpha^r(\psi^{r-1}(x)) \geq 0\}
    \end{equation}
    will render the set $\mathcal{C} = \bigcap_{k=1}^r \mathcal{C}^k$ forward invariant for system \eqref{eq:dyn_gen}.
\end{lemma}

\section{Path Planning for Taxiing Aircraft} \label{sec:pathplanning}
In this section, we describe our implementation of a path planning algorithm for taxiing aircraft, similar to the approach from \cite{zhang2020software}. We outline the necessary steps to represent an airport's taxiway network as a graph $\mathcal{G} = (\mathcal{V}, \mathcal{E})$, with nodes $\mathcal{V}$ indicating key positions such as intersections and gates, while edges $\mathcal{E}$ signify the connecting taxiways. To accurately capture the complexity of the taxiway network, all relevant nodes must defined with accurate latitude and longitude coordinates, including entry and exit points, intersections, and holding points that aircraft must navigate during taxiing.

\begin{figure}[h]
    \centering
    \includegraphics[width=\columnwidth]{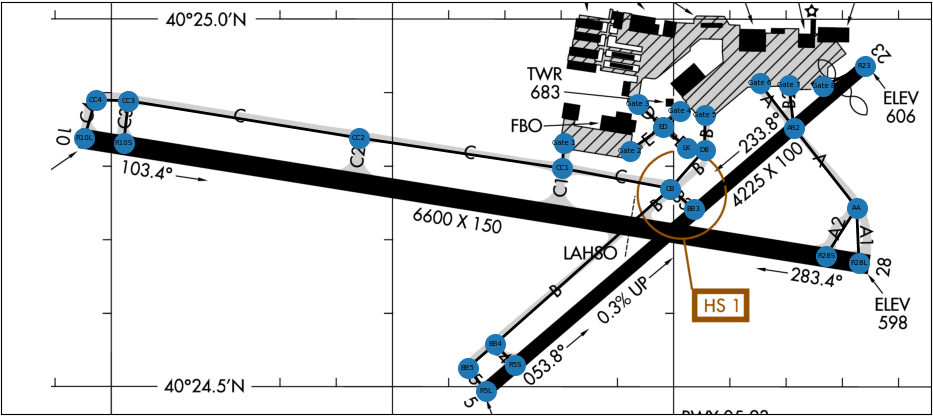}
    \caption{Undirected graph representation of Purdue University Airport (LAF) for autonomous aircraft navigation, illustrating runways and taxiways as nodes and edges for control module integration.}
    \label{fig:undirected_graph}
\end{figure}

We first construct an undirected graph of the airport that represents the bidirectional navigation of taxiways within an airport. 
In this graph, edges between nodes are weighted to reflect the actual distances between taxi points. For a visual example of such an undirected graph on a real airport, we show a potential mapping of the Purdue University Airport in Figure~\ref{fig:undirected_graph}. 

\begin{figure}[h]
    \centering
    \includegraphics[width=\columnwidth]{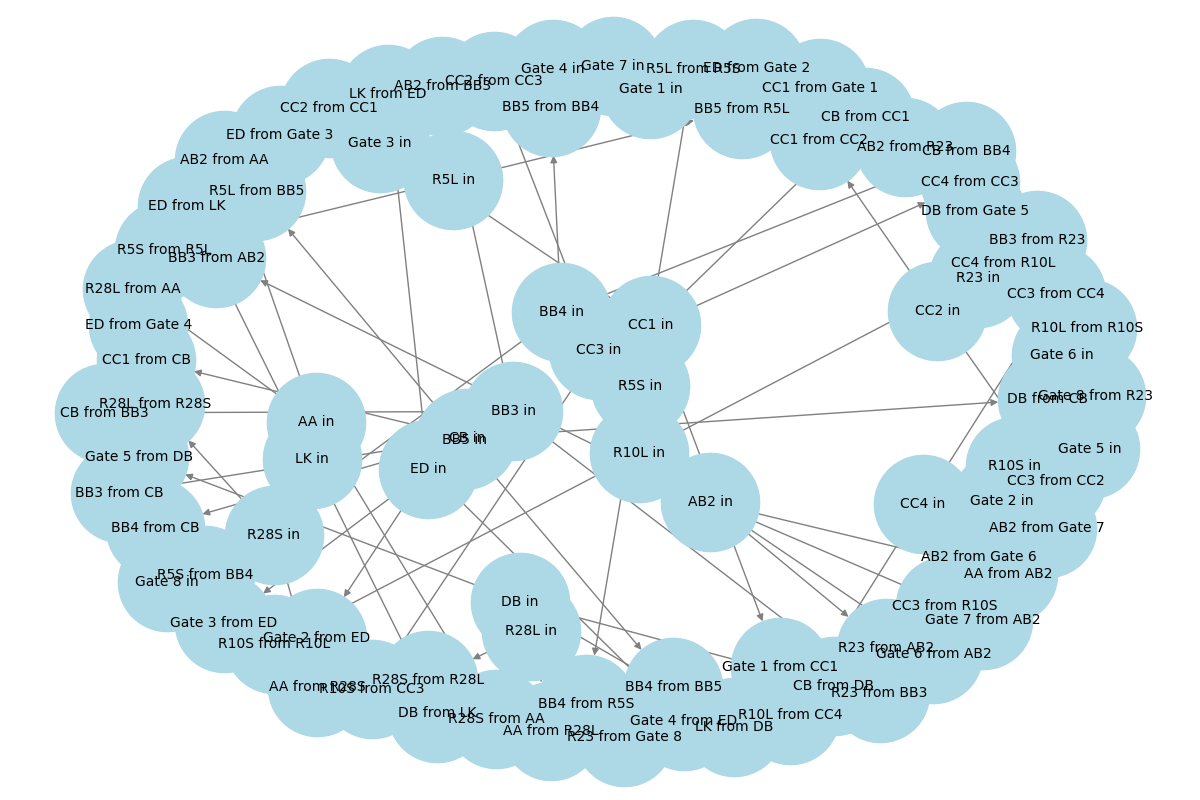}
    \caption{A directed graph representation of Purdue University Airport's taxiways and gates, with edges representing possible aircraft movement paths for autonomous navigation systems.}
    \label{fig:directed_graph}
\end{figure}

However, since an undirected graph does not accurately model the way an aircraft moves on the ground (e.g., a typical aircraft turning radius will not allow it to return directly to a node it just left), we need to use a directed graph to accurately reflect real-world airport operations, where certain taxiways may only be navigated in specific directions depending on the aircraft's recent location. This conversion process was guided by a method detailed in \cite{zhang2020software}, which facilitates the transformation of bidirectional taxiways into directed edges, ensuring that directional restrictions and operational constraints are properly captured. 
The resulting directed graph serves as a foundation for pathfinding, allowing us to apply Dijkstra's algorithm to calculate the shortest feasible route while adhering to directional taxiway limitations. See Figure~\ref{fig:directed_graph} for a visualization of the undirected graph mapping of the Purdue Airport in Figure~\ref{fig:undirected_graph} to a directed graph that accounts for the allowable origin and destination nodes of taxiing aircraft.

Using Dijkstra’s algorithm, we can identify the shortest path between any two nodes in the graph, which are represented as an ordered list of $n$ waypoints $\{d_1, \dots, d_n\}$, where $d_i \in \mathcal{V}, \forall i \in [n]$. This algorithm is particularly well-suited for this application due to its efficiency in computing the shortest path by considering the weights of the edges, which represent the distances between taxiway points. 

\subsection{Reference Trajectory Generation}
Once the shortest valid path has been selected, we must generate a reference trajectory from the set of sequential waypoints $\{d_1, \dots, d_n\}$ selected by the pathfinding algorithm from \cite{zhang2020software}. Given a desired target velocity $v_{\text{ref}} > 0$ and turning radius for the given aircraft $q > 0$ (i.e., the ideal turning rate of the aircraft were it to trace a circle of radius $q$), we generate a reference trajectory consisting of desired positions $p_{\text{ref}}(t) \in \mathbb{R}^2$ and orientations $\theta_{\text{ref}}(t) \in \mathbb{R}$ over the time interval $t \in [0, T]$, which may be concatenated into the reference state vector $x_{\text{ref}}(t) = [p_{\text{ref}}(t)^\top, \theta_{\text{ref}}(t)]^\top$, as follows. First, we replace all waypoints $d_i$ for $1 < i < n$ with a set of points that trace the arc of a circle with radius $q$ such that the start and end points of the arc are tangent to the angle generated by connecting the points $d_{i-1} \rightarrow d_i \rightarrow d_{i+1}$. We then generate legs linearly between points $p(t)$ such that 
\begin{equation}
    \frac{\Vert p_{\text{ref}}(t+dt) - p_{\text{ref}}(t) \Vert}{dt} = v_{\text{ref}},
\end{equation} 
where $dt$ is the chosen time resolution of the trajectory and $\theta_{\text{ref}}(t)$ is the angle that points from $p_{\text{ref}}(t)$ to $p_{\text{ref}}(t+dt)$ (i.e., $\theta_{\text{ref}}(t)$ is the angle of the line tangent to $p_{\text{ref}}(t)$).

\section{Safety Framework Using MPC-CBF} \label{sec:safety_framework}
In this section, we present a method for multi-obstacle avoidance and safe reference trajectory tracking using HOCBFs. To model the dynamics of the taxiing aircraft at a high level, we use simple 2D unicycle dynamics to model how the aircraft moves on the ground as follows,

\begin{equation}
    x = 
    \begin{bmatrix}
        p \\ v \\ \theta \\ \omega
    \end{bmatrix}
    , \;\; \dot{x} =
    \begin{bmatrix}
        v \\ a \\ \omega \\ \tau
    \end{bmatrix},
\end{equation}
where $p, v, a \in \mathbb{R}^2$ are the position, velocity, and acceleration and $\theta, \omega, \tau \in \mathbb{R}$ are the orientation, angular velocity, and angular acceleration of the aircraft, respectively. We apply control via acceleration with force and torque inputs $u = [u_F, u_{\tau}]^\top$, and control limits $u_F \in [F_{min}, F_{max}]$ and $u_\tau \in [\tau_{min}, \tau_{max}]$, where
\begin{equation}
    a(u_F) = \frac{u_F}{m}
    \begin{bmatrix}
        \cos \theta \\ \sin \theta
    \end{bmatrix}
    , \;\;
    \tau(u_\tau) = \frac{u_\tau}{I},
\end{equation}
and $m, I$ are the total mass and inertia of the aircraft, respectively. Under this formulation, we have a control affine system for the taxiing aircraft with dynamics given by
\begin{equation} \label{eq:unicycle_dyn}
    \dot{x} =
    \begin{bmatrix}
        v \\ - \frac{f_0}{m} v \\ \omega \\ 0
    \end{bmatrix}
    +
    \begin{bmatrix}
        0 & 0 \\
        0 & 0 \\
        \frac{\cos \theta}{m}  & 0 \\
        \frac{\sin \theta}{m} & 0 \\
        0 & 0 \\
        0 & \frac{1}{I}
    \end{bmatrix}
    \begin{bmatrix}
        u_F \\ u_\tau
    \end{bmatrix},
\end{equation}
where $f_0$ is the force applied by friction. 
Using these dynamics, we can now define safety conditions for collision avoidance and reference trajectory tracking using HOCBFs.

\subsection{Collision Avoidance} \label{sec:collision_avoid}
To mathematically formulate our goal of collision avoidance, we define the following barrier function candidate
\begin{equation} \label{eq:barrierfunc_obs}
    h_{o}(x, x_o) = \frac{1}{2}\left({\Vert p_o - p \Vert}^2 - \Delta^2 \right),
\end{equation}
where $\Delta>0$ is the desired safe distance we would like the aircraft to maintain from any detected obstacles. Since \eqref{eq:barrierfunc_obs} has a least relative degree of two, we need to compute the first and second derivatives of $h_o$, which are given by 

\begin{equation} \label{eq:barrierfunc_obs_dot}
    \dot{h}_{o}(x, x_o) = (p_o - p)^\top v
\end{equation}
and
\begin{equation} \label{eq:barrierfunc_obs_ddot}
    \ddot{h}_{o}(x, x_o, u) = (p_o - p)^\top a(u_F)  - \Vert v \Vert^2.
\end{equation}

\noindent
We use \eqref{eq:barrierfunc_obs} and \eqref{eq:barrierfunc_obs_dot} to compute the following functions

\begin{equation}\label{eq:second_order_BF_obs}
    \begin{aligned}
        \psi_{o}^0(x, x_o) &= h_o(x, x_o) \\
        \psi_{o}^1(x, x_o) &= \dot{\psi}_{o}^0(x, x_o) + \alpha^0_o\psi_{o}^0(x, x_o),
    \end{aligned}
\end{equation}
which we use to define the set $\mathcal{C}_o = \mathcal{C}^1_o \cap \mathcal{C}^2_o$ by \eqref{eq:HO_sets}. Therefore, by Lemma~\ref{lem:HOCBF}, the condition we must satisfy to guarantee that the aircraft will not collide with obstacle $o$, given a control law $u$, is 

\begin{equation}\label{eq:safety_cond_obs}
        \dot{\psi}_{o}^1(x, x_o, u) + \alpha^1_o\psi_{o}^1(x, x_o) \geq 0.
\end{equation}

Note that only the force control input $u_F$ appears in the second derivative of $h_o$.
Thus, 
any turning command $u_\tau$ must come from a nominal command signal.

\subsection{Safe Reference Tracking} \label{sec:safe_ref_tracking}
To mathematically formulate our goal of safe reference tracking, we define a barrier function candidate similar to \eqref{eq:barrierfunc_obs}. However, instead of trying to avoid a point by some safety radius $d$, we want to stay within some distance $w>0$ of the reference trajectory, for all time $t$, defined as follows

\begin{equation} \label{eq:barrierfunc_ref}
    h_{\text{ref}}(x, x_{\text{ref}}) = \frac{1}{2} \left(w^2 - \Vert p - p_{\text{ref}} \Vert^2 \right) , \forall t.
\end{equation}
Again, since \eqref{eq:barrierfunc_ref} has a least relative degree of two, we need to compute the first and second derivatives of $h_{\text{ref}}$, which are given by  
\begin{equation} \label{eq:barrierfunc_ref_dot}
    \begin{aligned}
        \dot{h}_{\text{ref}}(x, x_{\text{ref}}) 
        &= (p-p_{\text{ref}})^\top(v_{\text{ref}} - v)
    \end{aligned}
\end{equation}
and
\begin{equation} \label{eq:barrierfunc_ref_ddot}
    \begin{aligned}
        \ddot{h}_{\text{ref}}(x, x_{\text{ref}}, u) 
        &= (p -p_{\text{ref}})^\top (a_{\text{ref}} - a(u_F)) -\Vert v_{\text{ref}}-v \Vert^2. 
    \end{aligned}
\end{equation}

However, we see that, in addition to the difference in sign, \eqref{eq:barrierfunc_ref_dot} and \eqref{eq:barrierfunc_ref_ddot} differ most from \eqref{eq:barrierfunc_obs_dot} and \eqref{eq:barrierfunc_obs_ddot} in that $x_{\text{ref}}$ now is varying with time, causing $v_{\text{ref}}$ and $a_{\text{ref}}$ to appear in the derivative computations. If we let the commanded ground speed of the aircraft at time $t$ be define as
\begin{equation}
    s_{\text{ref}}(t) = \Vert v_{\text{ref}}(t) \Vert,
\end{equation}
then we can compute the acceleration of a given reference trajectory at time $t$ as
\begin{equation} \label{eq:a_ref}
    a_{\text{ref}}(t) =  \dot{s}_{\text{ref}}(t) \mathbf{T}(t) + \kappa(t) s_{\text{ref}}(t)^2 \mathbf{N}(t),
\end{equation}
where $\mathbf{T}(t)$ and $\mathbf{N}(t)$ are the unit vectors that are tangent and normal to the reference trajectory at time $t$, respectively, and $\kappa(t)$ is the curvature of the trajectory at time $t$. Note that if the commanded speed is constant, then \eqref{eq:a_ref} becomes
\begin{equation}
    a_{\text{ref}}(t) = \kappa(t) s_{\text{ref}}(t)^2 \mathbf{N}(t).
\end{equation}
Further, note that we may compute the unit vector normal to the trajectory using $\theta_{\text{ref}}(t)$, which is defined to be the angle tangent to $p_{\text{ref}}(t), \forall t$. Similar to Section~\ref{sec:collision_avoid}, we compute the following functions
\begin{equation}\label{eq:second_order_BF_ref}
    \begin{aligned}
        \psi_{\text{ref}}^0(x, x_{\text{ref}}) &= h_{\text{ref}}(x, x_{\text{ref}}) \\
        \psi_{\text{ref}}^1(x, x_{\text{ref}}) &= \dot{\psi}_{\text{ref}}^0(x, x_{\text{ref}}) + \alpha^0_{\text{ref}}(\psi_{\text{ref}}^0(x, x_{\text{ref}})),
    \end{aligned}
\end{equation}
which define the time-varying safety requirement set $\mathcal{C}_{\text{ref}}(t) = \mathcal{C}^1_{\text{ref}}(t) \cap \mathcal{C}^2_{\text{ref}}(t)$. Thus, by Lemma~\ref{lem:HOCBF}, the second condition that we must satisfy to guarantee that the aircraft does not exceed a distance of $w>0$ away from the reference trajectory, where $w$ may be half of the width of the runway, is
\begin{equation}\label{eq:first_order_safety_cond}
        \dot{\psi}_{\text{ref}}^1(x(t), x_{\text{ref}}(t), u) + \alpha^1_{\text{ref}}(\psi_{\text{ref}}^1(x(t), x_{\text{ref}}(t))) \geq 0, \forall t.
\end{equation}

\subsection{MPC-CBF Framework}
Using the conditions for obstacle avoidance and safe reference tracking defined in the previous subsections, we are prepared to formalize the MPC-CBF problem as follows. Given a prediction horizon of $N \in \mathbb{N}$, we minimize the following cost function constrained by the system dynamics in \eqref{eq:unicycle_dyn} and our two safety requirements
\begin{equation} \label{eq:MPC-CBF}
    \begin{aligned}
        \min_{u \in \mathcal{U}} & \sum_{k = 1}^{N} (x(k)- x_{\text{ref}}(k))^\top Q (x(k)- x_{\text{ref}}(k)) + u(k)^\top R u(k) \\ 
        \text{s.t.} & \quad x(k+1) = f(x(k)) + g(x(k))u(k) \\
        & \quad \dot{\psi}_{\text{ref}}^1(x(k), x_{\text{ref}}(k), u(k)) + \alpha^1_{\text{ref}}\psi_{\text{ref}}^1(x(k), x_{\text{ref}}(k)) \geq 0 \\
        & \quad \dot{\psi}_{o}^1(x(k), x_o, u(k)) + \alpha^1_o\psi_{o}^1(x(k), x_o) \geq 0 \\
        & \quad x(k) \in \mathcal{X}, \; u(k) \in \mathcal{U} ; \forall k \in [N],
    \end{aligned}
\end{equation}
where $Q \in \mathbb{R}^{n \times n}_{\geq 0}$ and $R \in \mathbb{R}^{m\times m}_{\geq 0}$ are cost function parameter matrices for reference tracking and control minimization, respectively. Note that the discrete-time nature of model predictive control requires that the constraint of the system dynamics also be approximated in discrete time. For this application, we use the Rung-Kutta method to approximate a zero-hold control law $u(k)$ over a specified time interval $dt>0$, for each step of the prediction horizon $k \in [N]$. Further, note that the use of a CBF safety condition for safe reference tracking may be viewed in a similar spirit to that of tube MPC methods \cite{morato2020model}; however, a full analysis and comparison of the two methods is left as a topic for future work.

\subsection{PID-CBF Framework}

To compare the performance of our model-based framework, 
we also designed a PID control scheme as follows. Given a discrete time step, \( dt > 0 \), and a reference signal, $x_{\text{ref}}(t) = [p_{\text{ref}}(t)^\top, \theta_{\text{ref}}(t)]^\top$, we aim to regulate the system's position and orientation through force and torque control inputs. To compute the error signals that feedback into both the force and torque inputs, we define the unit vector with angle $\theta_{\text{ref}}$ as
\begin{equation}
    p_\theta = \begin{bmatrix} \cos(\theta_{\text{ref}}) \\ \sin(\theta_{\text{ref}}) \end{bmatrix},
\end{equation}
and the vector pointing from $p_{\text{ref}}(t)$ to the position of the aircraft $p(t)$ at time $t$ as
\begin{equation}
    p_{\text{diff}}(t) = p(t) - p_{\text{ref}}(t).
\end{equation}

The error signal used to control the force input is computed as
\begin{equation}
    e_F = K_{s}(s_{\text{ref}} - \Vert v \Vert) + K_{a}(p_{\text{diff}} \cdot p_{\theta}),
\end{equation}
where $K_{s}>0$ scales the error between the current vehicle speed and the command speed and $K_{a}>0$ scales the error in tracking the reference trajectory, correcting position undershoot or overshoot.
A classical PI controller then gives the input for force as
\begin{equation}\label{eq:uF_PID}
    u_F = K_{pF} e_F + K_{iF} \int e_F \, dt,
\end{equation}
where $K_{pF},K_{iF}>0$ are parameters that tune the proportional and integral components of the PI controller, respectively.

The error signal used to control the torque input is computed as
\begin{equation}
    e_T = K_{al}(\theta_{\text{ref}} - \theta) - K_{t}(-p_{\text{diff}} \times p_\theta),
\end{equation}
where $K_{al}$ scales the error between the current orientation of the aircraft and the orientation of the reference trajectory (i.e., the orientation of the runway), and $K_{t}$ scales the signal for the aircraft to turn toward the reference trajectory. A classical PD controller then gives the input for torque as

\begin{equation}\label{eq:utau_PID}
    u_\tau = K_{pT} e_T + K_{dT} \left( \frac{de_T}{dt} \right),
\end{equation}
where $K_{pT},K_{dT}>0$ are parameters that tune the proportional and derivative components of the PD controller, respectively.

We can then use our safety conditions defined in Sections~\ref{sec:collision_avoid} and \ref{sec:safe_ref_tracking} to filter the control signal $u_{\text{PID}} = [u_F, u_\tau]^\top$, as computed by \eqref{eq:uF_PID} and \eqref{eq:utau_PID}, using the following quadratic program.

\begin{equation} \label{eq:PID-CBF}
    \begin{aligned}
        \min_{u \in \mathcal{U}} & \quad \frac{1}{2}\Vert u - u_{\text{PID}} \Vert^2 \\ 
        \text{s.t.}
        & \quad \dot{\psi}_{\text{ref}}^1(x, x_{\text{ref}}, u) + \alpha^1_{\text{ref}}\psi_{\text{ref}}^1(x, x_{\text{ref}}) \geq 0 \\
        & \quad \dot{\psi}_{o}^1(x, x_o, u) + \alpha^1_o\psi_{o}^1(x, x_o) \geq 0.
    \end{aligned}
\end{equation}

\section{Simulations} \label{sec:simulations}
In this section, we compare the performance of our model-based MPC-CBF framework with the 
PID-CBF framework through simulation using the Purdue University Airport as an example. We first construct an undirected graph model of the airport by placing nodes at each intersection of the taxi runway, then we construct the directed graph and find the shortest path between hangars and runways using the path-planning algorithm described from \cite{zhang2020software} and generate the reference trajectory as described in Section~\ref{sec:pathplanning}. We then test both the MPC-CBF and PID-CBF frameworks' ability to track the reference trajectory under a variety of conditions
In the following subsections, we describe the parameterization of the simulations and control frameworks, then discuss the simulation results.

\subsection{Parameterization}
For all simulations, we set control limits $u_F \in [0,4]$ and $u_\tau \in [-10,10]$, and the parameter values $m=1$, $I_0=1$, $\Delta = 10$, and $w=8$  for the mass, inertia, obstacle safety radius, and reference tracking radius, respectively. Further, we set $\alpha^0_{\text{ref}} = \alpha^1_{\text{ref}} = \alpha^0_{o}=\alpha^1_{o} = 10$ for both frameworks.
The PID-CBF controller is instantiated with the following parameters: $K_{s} = 1$, $K_a = 0.001$, $K_{pF} = 1$,
$K_{iF} = 0.5$, $K_{dF} = 0$, $K_{al}=1$, $K_{t}=0.01$, $K_{pT} = 0.001$, $K_{iT} = 0$, $K_{dT} = 0.22$. The MPC-CBF controller is instantiated with the parameters $Q = 10^3 \times I^{n\times n}$ and $R = 10^3 \times I^{m \times m}$.

\subsection{Simulation Results}
We then test both the MPC-CBF and PID-CBF frameworks' ability to track the reference trajectory under the following conditions:
\begin{itemize}
    \item No external disturbances or obstacles
    \item No external disturbances with obstacles
    \item Constant crosswind with no obstacles
    \item Constant crosswind with obstacles 
\end{itemize}

\begin{figure}
    \centering
    \includegraphics[width=.9\columnwidth]{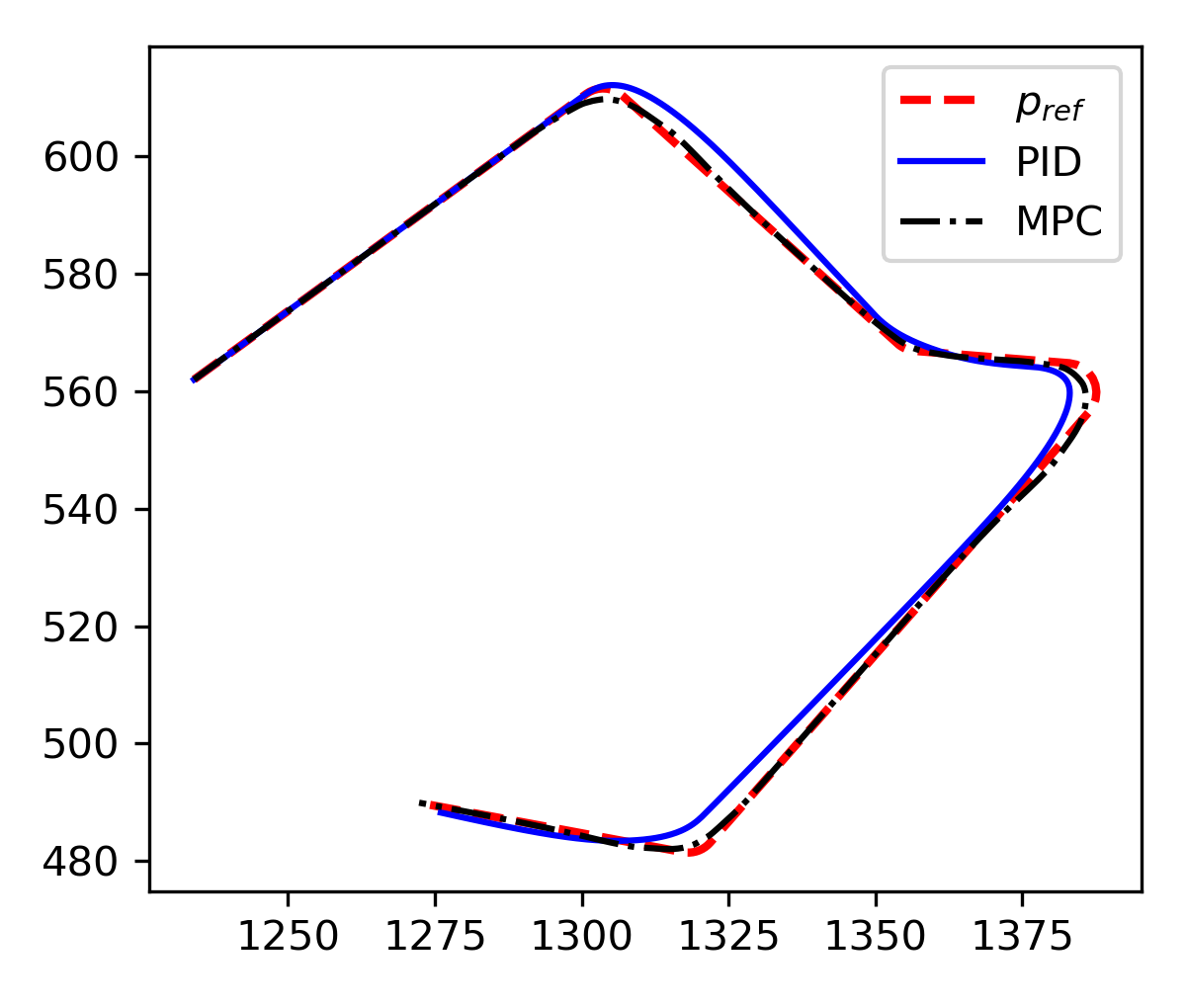}
    \caption{Comparison of the performance for the PID controller (blue solid) with the MPC controller (black dotted) in tracking the runway reference trajectory (red dashed) without external disturbances or obstacles. Note that the MPC controller outperforms the tuned PID controller at tracking the reference trajectory over time.}
    \label{fig:mpc-pid-compare}
\end{figure}

After some tuning, we achieved acceptable reference tracking with the PID-CBF framework with no external disturbances or obstacles, which we compare with the performance of the MPC-CBF framework under the same conditions in Figure~\ref{fig:mpc-pid-compare}, 
where we see that the MPC controller outperforms the tuned PID controller at tracking the reference trajectory over time. 

However, as is common with many PID-based controllers, the PID-CBF framework proved to be sensitive to both external disturbances and obstacles, causing it to fail all other tests. It is possible that better results could be achieved with additional tuning of the PID system parameters, which may be desirable in situations where the lower computational cost of the PID-CBF framework may outweigh the potential for instability; however, we did not perform such tuning in this work.

\begin{figure}
    \centering
    \includegraphics[width=.9\columnwidth]{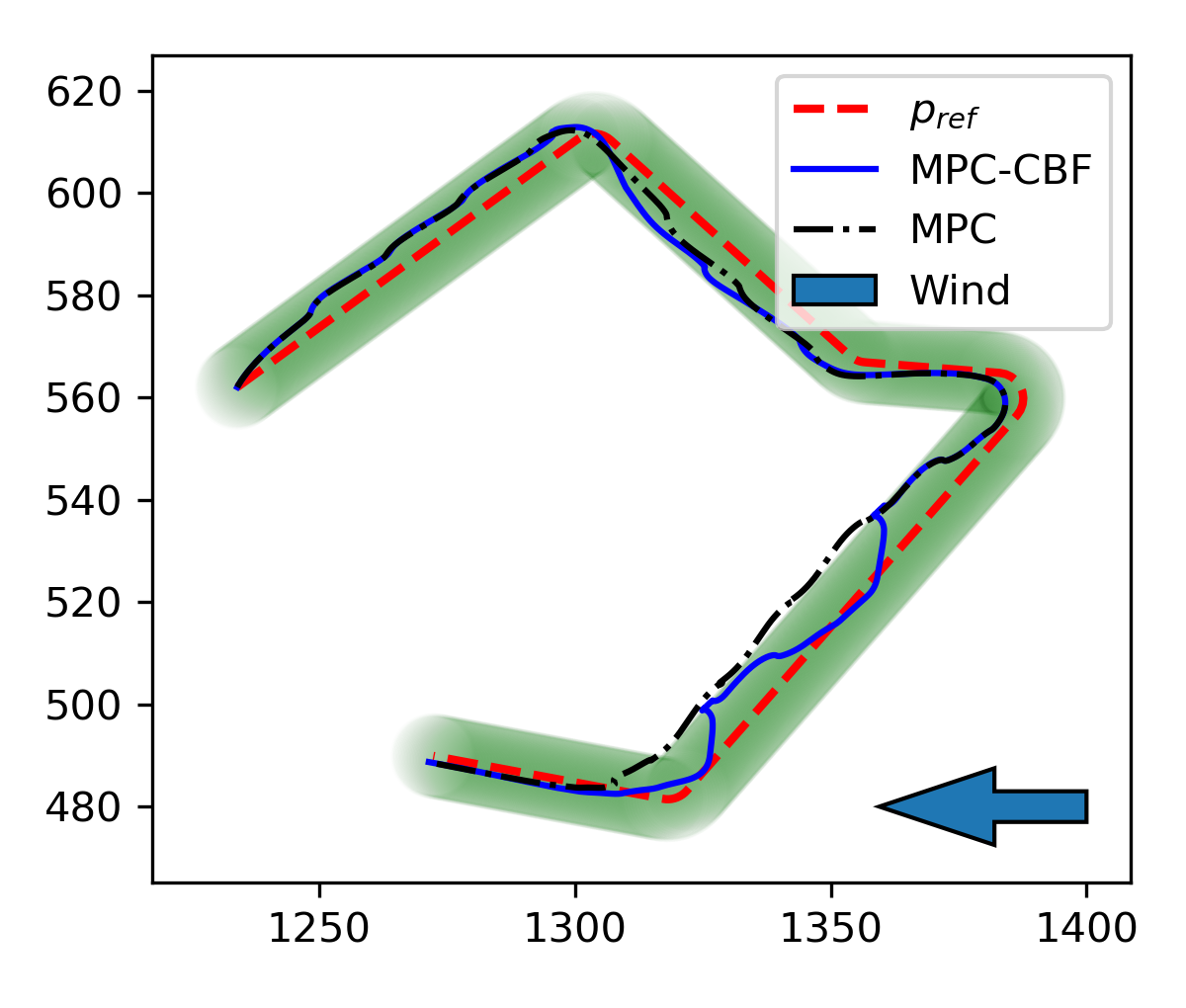}
    \caption{An example of the MPC-CBF framework (blue solid) under the external disturbance of a constant crosswind compared with the MPC controller without safe reference tracking (black dotted). Note that, without the reference tracking CBF active, the system is pushed outside of the defined safe region for the reference trajectory.}
    \label{fig:mpc-cbf-crosswind-compare}
\end{figure}

\begin{figure}
    \centering
    \includegraphics[width=.9\columnwidth]{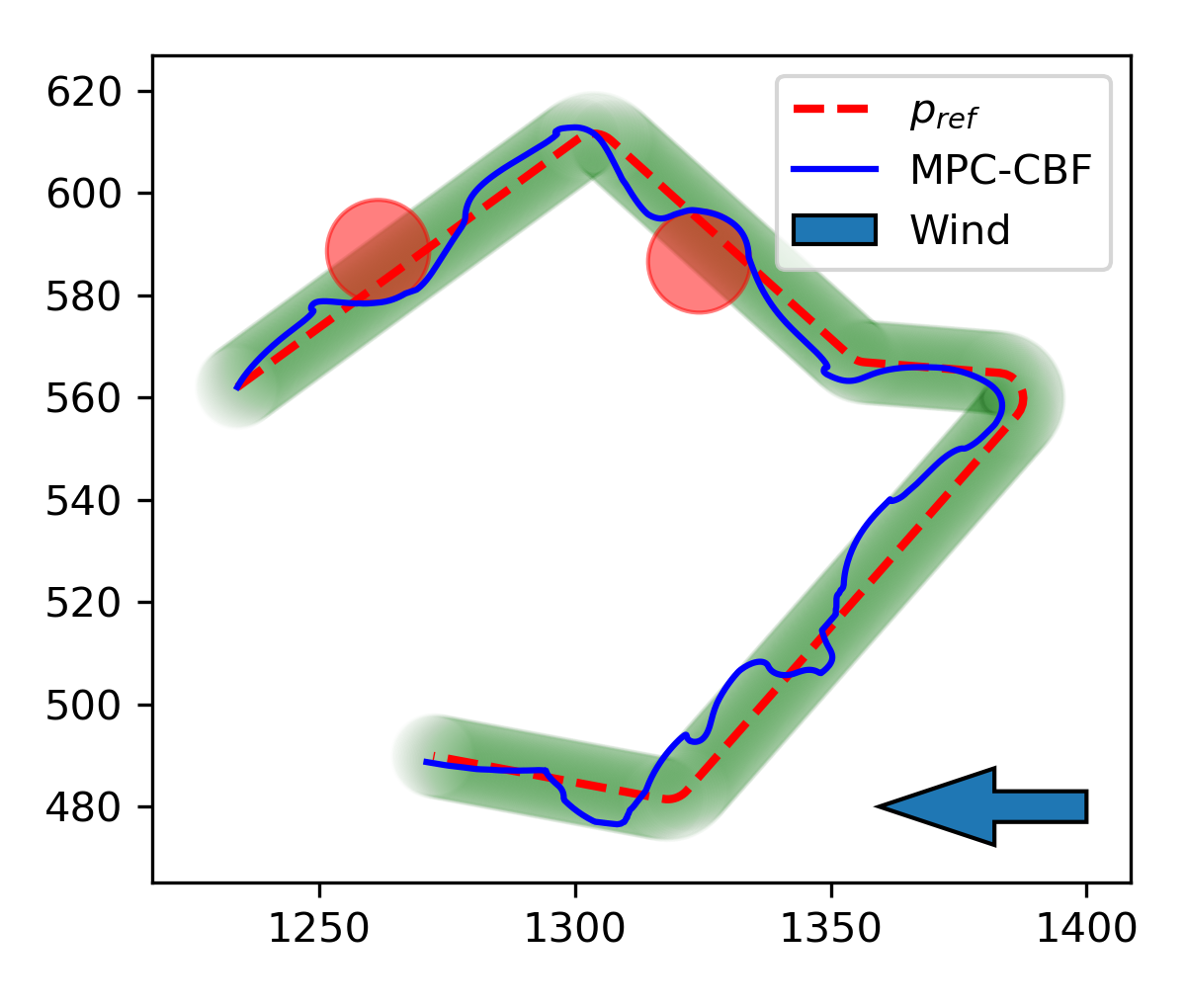}
    \caption{An example of the MPC-CBF framework controlled aircraft (blue solid) under the external disturbance of a constant crosswind while avoiding obstacles placed along the reference trajectory (red dashed). Note that, even under external disturbances, the MPC-CBF controller is capable of satisfying both the obstacle avoidance and safe reference tracking conditions for all time.}
    \label{fig:mpc-cbf-crosswind-obs}
\end{figure}

For the proposed MPC-CBF framework, we see that the controller is capable of handling both external disturbances, such as an unknown constant crosswind force, and obstacle collision avoidance, at the price of increased computation cost. We show in Figure~\ref{fig:mpc-cbf-crosswind-compare} how the safe reference tracking CBF assists the MPC controller to stay within the defined radius of the runway, whereas, without the CBF framework active, we see the MPC controller is forced outside of the defined safe region. Finally, we show in Figure~\ref{fig:mpc-cbf-crosswind-obs} the performance of the MPC-CBF framework in avoiding obstacles on the reference trajectory while experiencing a constant crossing force. Thus, we see that incorporating CBF safety constraints into the MPC framework increases the robustness of the taxiing system.

\section{Conclusion}
In this paper, we use a graph-based path planning algorithm for taxiing aircraft from \cite{zhang2020software} to generate a desired reference trajectory that should navigate an aircraft autonomously to and from takeoff. We define safety conditions for both obstacle avoidance and safe reference tracking using HOCBFs and use these conditions to define an MPC-CBF framework. We compare the performance of our MPC-CBF framework with a model-free PID-CBF framework and find that our MPC-CBF framework is capable of handling external disturbances while maintaining both safety conditions of obstacle avoidance and safe reference tracking, whereas the PID-CBF is unable to maintain safety conditions due to the sensitivity of the controller. Future directions of work include performing analysis on the formal guarantees provided by the MPC-CBF framework while accounting for the discrete-time nature of the MPC prediction horizon, as well as implementing our MPC-CBF framework on a real-world autonomous fixed-wing aircraft.   

\normalem
\bibliographystyle{IEEEtran}
\bibliography{references}

\end{document}